\documentclass[aps,prl,twocolumn,superscriptaddress,floatfix]{revtex4}
\usepackage{latexsym,hyperref,color,amssymb,graphicx}
\newcommand{\beq}{\begin{equation}}
\newcommand{\eeq}{\end{equation}}
\newcommand{\bea}{\begin{eqnarray}}
\newcommand{\eea}{\end{eqnarray}}
\newcommand{\bean}{\begin{eqnarray*}}
\newcommand{\eean}{\end{eqnarray*}}
\newcommand{\bei}{\begin{itemize}}
\newcommand{\eei}{\end{itemize}}
\newcommand{\ben}{\begin{enumeration}}
\newcommand{\een}{\end{enumeration}}

\newcommand{\thalf}{\textstyle{ {1 \over 2} } }

\definecolor{darkorange}{rgb}{.6,.2,.0}

\definecolor{darkgreen}{rgb}{0.0,0.7,0.0}

\begin{document}

\title{Rydberg-London Potential for Diatomic Molecules
and Unbonded Atom Pairs}

\author{Kevin Cahill}
\email{cahill@unm.edu}
\affiliation{Department of Physics and Astronomy, 
University of New Mexico, Albuquerque, NM 87131}
\affiliation{Center for Molecular Modeling, 
Center for Information Technology, 
National Institutes of Health, 
Bethesda, Maryland 20892-5624}
\author{V.~Adrian Parsegian}
\email{aparsegi@helix.nih.gov}
\affiliation{National Institute of Child Health
and Human Development, National Institutes of Health, 
Bethesda, Maryland 20892-0924}

\date{\today}

\begin{abstract}
We propose and test a pair potential that
is accurate at all relevant distances
and simple enough 
for use in large-scale computer simulations.  
A combination of the Rydberg potential from spectroscopy 
and the London inverse-sixth-power energy, 
the proposed form fits 
spectroscopically determined potentials
better than the Morse, Varnshi,
and Hulburt-Hirschfelder potentials
and much better than the 
Lennard-Jones and harmonic potentials.
At long distances, it 
goes smoothly to the correct London force 
appropriate for gases and preserves 
van der Waals's ``continuity of the gas and liquid states," 
which is routinely violated by coefficients assigned 
to the Lennard-Jones 6-12 form.
\end{abstract}

%\pacs{82.35.Pq,34.20.Cf,77.84.Jd}

\maketitle
There are at least three classes of 
interatomic potentials.
Commercial codes
use the Lennard-Jones 
\beq
V_{LJ}(r) = |V(r_0)| \, \left[ \left( \frac{r_0}{r} \right)^{12}
- 2 \, \left( \frac{r_0}{r} \right)^6 \right]
\label{LJ}
\eeq
and harmonic 
\beq
V_H(r) = V(r_0) 
+ \frac{(r - r_0)^2}{2} \, 
\frac{d^2 V(r_0)}{dr^2} 
\label{Ha}
\eeq
forms, which
are accurate
near the minimum at \(r=r_0\)\@.
But this first class of potentials may be too simple
for complex materials
away from from equilibrium.
\par
The Morse~\cite{Morse1929}
\beq
V_M(r) = |V(r_0)| \, \left[
\left( 1 - e^{-\kappa \, x} \right)^2
- 1 \right]
\label{Morse}
\eeq
(\(x = r - r_0\), \( \kappa = \sqrt{k_e/(2|V(r_0)|)} \)),
Varnshi~\cite{Varnshi1957}
\beq
V_V(r) = |V(r_0)| \, \left[
\left( 1 - \frac{r_0}{r} \, e^{- \beta \, (r^2 - r_0^2)} 
\right)^2
- 1 \right],
\label{Varnshi}
\eeq
(\( \beta = (\kappa r_0^2 -1)/(2r_0^2) \)), 
and Hulburt-Hirschfelder~\cite{Hulbert1961}
\beq
V_{HH}(r) = V_M(r) + |V(r_0)| \, 
c'\kappa^3x^3\,e^{-2\kappa x}(1+\kappa b'x)
\label{HH}
\eeq
(for \(b'\), \(c'\), see ref.\cite{Vanderslice1961})
potentials represent
a second class of potentials~\cite{Vanderslice1961}
accurate over a wider
range of distances.
\par
Quantum 
chemists~\cite{Scoles1984,Aziz1993,Meath2001,Tang2003,Varandas2004}
have derived a third class 
that reproduce spectroscopic
and thermodynamic data with impressive fidelity.
But the potentials of this class
involve many parameters and
may be too cumbersome for use in large-scale simulations.
\par
We propose and test a form
\beq
V(r) = a e^{-b \, r}\,( 1 - c \, r ) 
- \frac{d}{r^6 + e \, r^{-6}}
\label{phenpot}
\eeq
that is nearly as accurate as the class-3 potentials 
but simpler than many class-2 potentials.
It is a combination of the Rydberg formula 
used in spectroscopy 
and the London formula for pairs of atoms.
In Eq.(\ref{phenpot}), the terms involving  \(a\), \(b\), and \(c\) 
were proposed by Rydberg~\cite{Rydberg1931}
to incorporate spectroscopic data,
but were largely ignored until 
recently~\cite{Ferrante1991,Murrell1984}.
The constant \(d = C_6\)
is the coefficient of the London tail.
The new term \( e \, r^{-6} \) cures
the London singularity.
As \(r \to 0\),
\( V(r) \to a\), finite;
as \(r \to \infty\), \(V(r)\) 
approaches the London term,
\(V(r) \to - d/r^6 = -C_6/r^6\)\@.
In a perturbative analysis~\cite{Cahill2003},
the \(a, b, c\) terms 
arise in first order, and the \(d\) term
in second order. 
\par
\begin{figure}
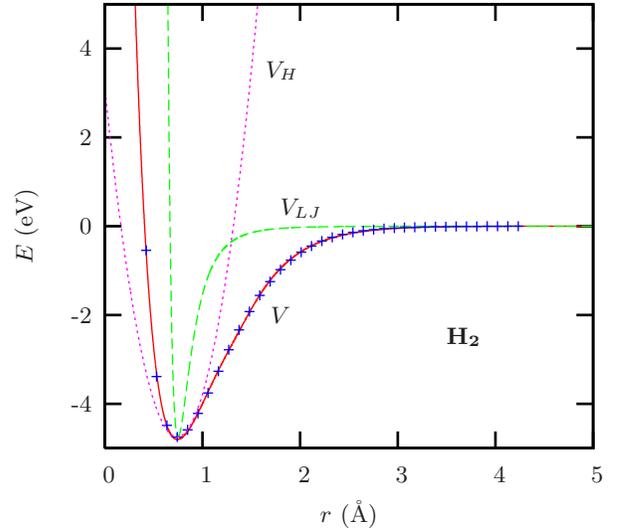

\centering
\input RKRH2
\caption{The hybrid form \(V\)
(with \(a = 53.8\) eV,
\(b = 2.99\) \AA\(^{-1}\),
\(c = 2.453\) \AA\(^{-1}\),
\(d = C_6 = 3.884\) eV\,\AA\(^6\),
and \(e = 47.6\) \AA\(^{12}\))
(solid, red) 
fits the RKR spectral points for 
the ground state of
molecular hydrogen (pluses, blue)
and gives the correct London tail for \(r > 3\) \AA\@.
The Lennard-Jones \(V_{LJ}\) (dashes, green) and harmonic
\(V_H\) (dots, magenta) forms fit
only near the minimum.}
\label{rkr}
\end{figure}
\begin{figure}
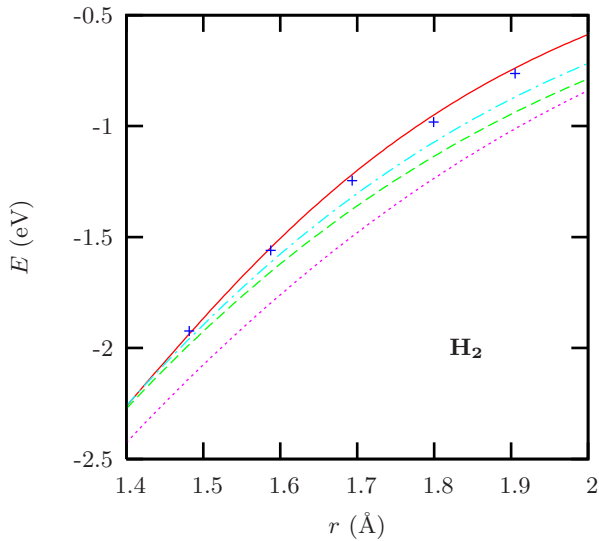

\centering
\input H2REV
\caption{The hybrid form \(V\)
(Eq.(\ref{phenpot}) as in Fig.~\ref{rkr},
solid, red) 
fits the RKR spectral points for 
the ground state of
molecular hydrogen (pluses, blue).
The Morse \(V_{M}\) (Eq.(\ref{Morse}), dashes, green),
Varnshi \(V_V\) (Eq.(\ref{Varnshi}), dots, magenta), and 
Hulburt-Hirschfelder
\(V_{HH}\) (Eq.(\ref{HH}), dot-dash, cyan) 
are too low for
\(1.5 < r < 3\) \AA\@.}
\label{h2rev}
\end{figure}
\begin{figure}
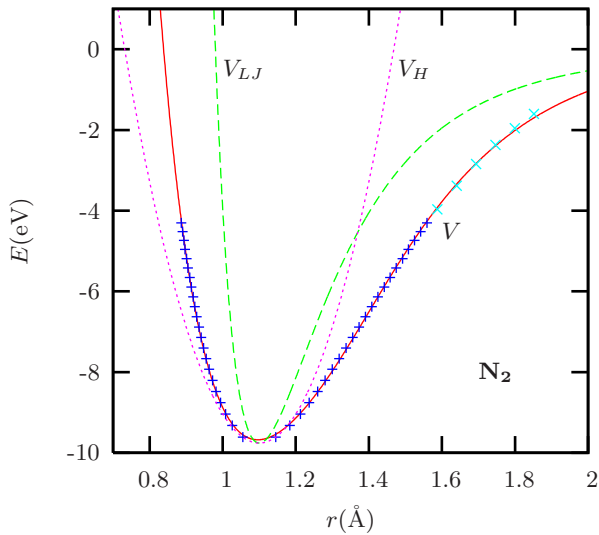

\centering
\input RKRN2
\caption{For the ground state of molecular nitrogen,
the proposed potential \(V\) of Eq.(\ref{phenpot}) 
(with \(a = 3.25\) keV,
\(b = 4.30\) \AA\(^{-1}\),
\(c = 1.1906\) \AA\(^{-1}\),
\(d = 14.1\) eV\,\AA\(^6\), and
\(e = 27.1\) \AA\(^{12}\))
(solid, red) 
fits the RKR spectral points (pluses, blue)
and the FO points (x's, cyan)\@.
\(V_{LJ}\) and \(V_H\) as in Fig.~\ref{rkr}\@.}
\label{rkrn2}
\end{figure}
\begin{figure}
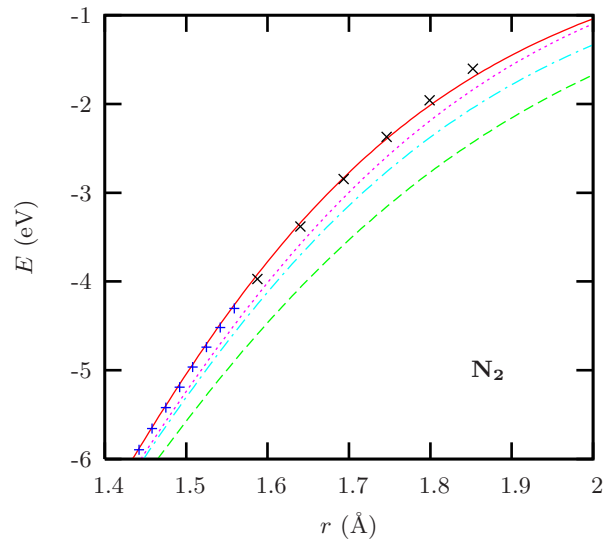

\centering
\input N2REV
\caption{For the ground state of molecular nitrogen,
the hybrid form \(V\) (Eq.(\ref{phenpot}),
as in Fig.~\ref{rkrn2}, solid, red) 
fits the RKR spectral points (pluses, blue)
and the FO points (x's, black)\@.
\(V_{M}\), \(V_V\), \&
\(V_{HH}\) as in Fig.~\ref{h2rev}.}
\label{n2rev}
\end{figure}
\begin{figure}
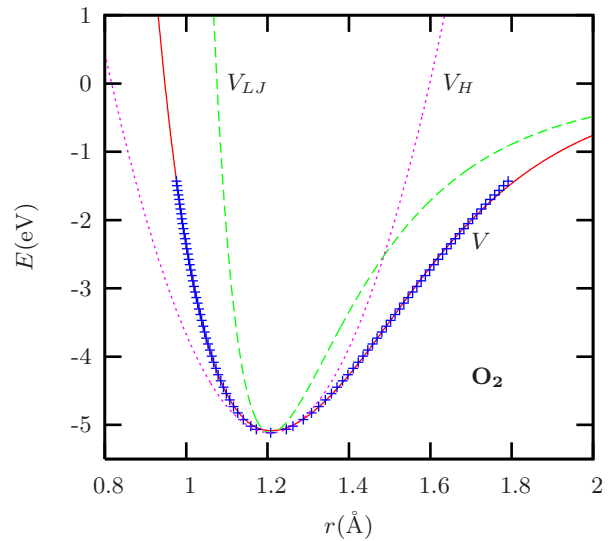

\centering
\input RKRO2
\caption{For the ground state of molecular oxygen,
the potential \(V\) of Eq.(\ref{phenpot}) 
(with \(a = 3.61\) keV,
\(b = 4.48\) \AA\(^{-1}\),
\(c = 1.05\) \AA\(^{-1}\),
\(d = 16.08\) eV\,\AA\(^6\), and
\(e = 58.4\) \AA\(^{12}\))
(solid, red) fits
the RKR points (pluses, blue)\@.
\(V_{LJ}\) and \(V_H\) as in Figs.~\ref{rkr} 
\& \ref{rkrn2}\@.}
\label{rkro2}
\end{figure}
\begin{figure}
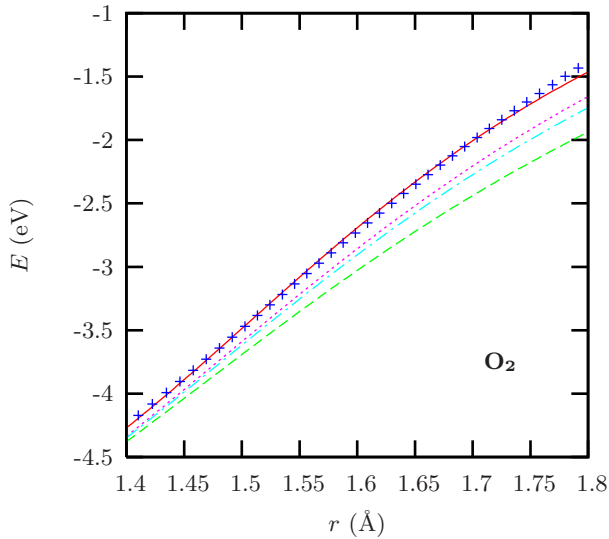

\centering
\input O2REV
\caption{For the ground state of molecular oxygen,
the hybrid form \(V\) (Eq.(\ref{phenpot}),
as in Fig.~\ref{rkro2}, solid, red) 
fits the RKR spectral points (pluses, blue)\@.
\(V_{M}\), \(V_V\), \&
\(V_{HH}\) as in Fig.~\ref{h2rev}.}
\label{o2rev}
\end{figure}
\begin{figure}
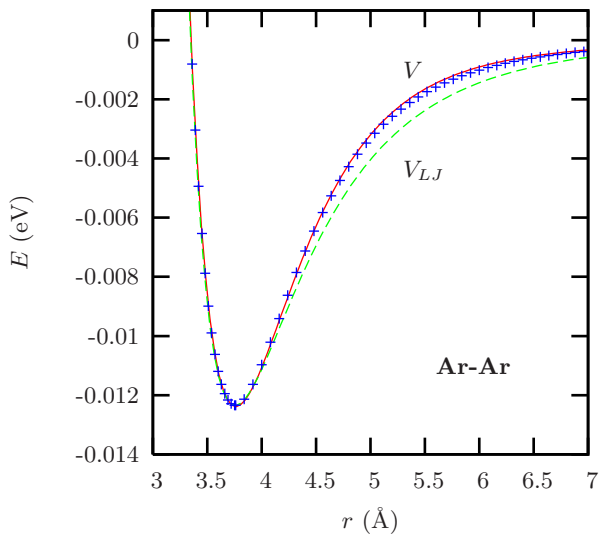

\centering
\input AR2374
\caption{The potential \(V\) (Eq.(\ref{phenpot}) 
for the Ar-Ar ground state  
with \(a = 1720\) eV,
\(b = 2.6920\) \AA\(^{-1}\),
\(c = 0.2631\) \AA\(^{-1}\),
\(d = 37.943\) eV\,\AA\(^6\), and
\(e = 177588\) \AA\(^{12}\))
(solid, red) fits
the Ar\(_2\) Aziz potential (pluses, blue)
with the correct London tail.
When matched at the minimum,
the Lennard-Jones form \(V_{LJ}\) 
(Eq.(\ref{LJ}) with \(r_0 = 3.757\) \AA\ 
and \(V(r_0) = -0.01234\) eV) (dashes, green) 
is too low for \(r > 4\) \AA.}
\label{ar237}
\end{figure}
\begin{figure}
\centering
\input AR24
\caption{Positive potential \(V\) (Eq.(\ref{phenpot}) 
as in Fig.~\ref{ar237}),
Ar\(_2\) Aziz potential (pluses, blue),
L-J form \(V_{LJ}\) (Eq.(\ref{LJ})
as in Fig.~\ref{ar237})\@.}
\label{ar2}
\end{figure}
\begin{figure}
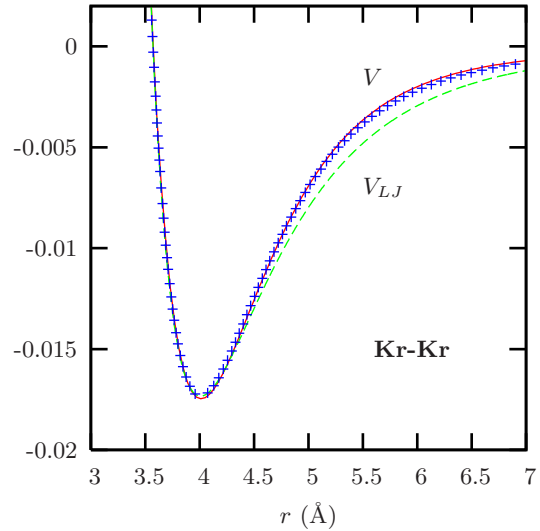

\centering
\input KR2374
\caption{The potential \(V\) (Eq.(\ref{phenpot})
for the Kr-Kr ground state  
with \(a = 2499\) eV,
\(b = 2.5249\) \AA\(^{-1}\),
\(c = 0.2466\) \AA\(^{-1}\),
\(d = 78.214\) eV\,\AA\(^6\), and
\(e = 199064\) \AA\(^{12}\))
(solid, red) fits
the Kr\(_2\) Aziz points (pluses, blue)
with the correct London tail.
When matched at the minimum,
the Lennard-Jones form \(V_{LJ}\) 
(Eq.(\ref{LJ}) with \(r_0 = 4.008\) \AA\ 
and \(V(r_0) = -0.017338\) eV) (dashes, green) 
is too low for \(r > 4\) \AA.}
\label{kr237}
\end{figure}
\begin{figure}
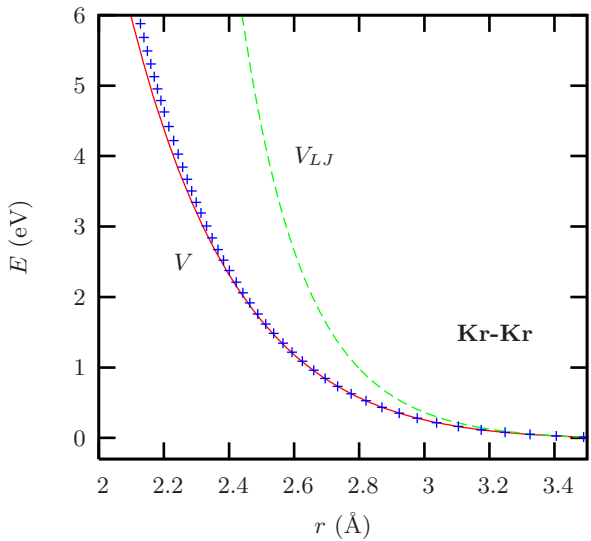

\centering
\input KR24
\caption{Positive potential \(V\) (Eq.(\ref{phenpot}) 
as in Fig.~\ref{kr237}),
Kr\(_2\) Aziz points (pluses, blue),
L-J form \(V_{LJ}\) (Eq.(\ref{LJ})
as in Fig.~\ref{kr237})\@.}
\label{kr2}
\end{figure}
\par
To test whether the 
hybrid \(V(r)\)
can represent covalent bonds
far from equilibrium,
we used Gnuplot~\cite{Gnuplot}
to fit Eq.(\ref{phenpot}) 
to empirical potentials for % Remark: we had 'of,' not 'for.'
molecular H\(_2\), N\(_2\), and O\(_2\)
obtained from spectroscopic 
data~\cite{Weissman1963,Krupenie1977,Krupenie1972}
by the RKR (Rydberg~\cite{Rydberg1931}, Klein~\cite{Klein1932},
Rees~\cite{Rees1947}) method,
setting \(d\) equal to the London values.
Figure~\ref{rkr} shows
that the hybrid potential of Eq.(\ref{phenpot}) (solid, red)
goes through the RKR points for H\(_2\) (pluses, blue)
from 0.5 to 4 \AA\@.
Fitted to the minimum,
the harmonic potential (\ref{Ha}) 
(dashes, green)
and the Lennard-Jones potential (\ref{LJ})
(dashes, blue)
are accurate only
near the minimum at 0.74 \AA\@.
Figures~\ref{rkrn2} \& \ref{rkro2}
show that the hybrid potential \(V(r)\)
fits the N\(_2\) and O\(_2\) 
RKR points~\cite{Krupenie1977,Krupenie1972}
and the N\(_2\) first-order (FO)
points~\cite{Ermler1982}
between 1 and 1.8 \AA\@.
The Lennard-Jones and harmonic potentials
of Eqs.(\ref{LJ}--\ref{Ha}) fit only near the minima.
\par
How does the hybrid form compare
with the class-2 potentials
of Eqs.(\ref{Morse}--\ref{HH})?
Figures~\ref{h2rev}, \ref{n2rev}, \&
\ref{o2rev} show that for 
\( 1.4 < r < 2 \) \AA\ 
the hybrid \( V(r) \)
is closer than (\ref{Morse}--\ref{HH})
to the RKR points.
A useful estimate of how well a 
particular potential \( V_P(r) \)
fits \(N\) data points \( V_D(r_i) \)
is the dimensionless error 
\beq
\delta = \frac{1}{|V_D(r_0)|} \,
\left[ \frac{1}{N} \,
\sum_{i = 1}^N 
\left( V_P(r_i) - V_D(r_i) \right)^2 \,
\right]^{1/2}.
\label{delta}
\eeq
For \(H_2\),
\(N_2\), and \(O_2\), the average
error \(\delta\) was 59.7 for \(V_H\),
49.3 for \(V_{LJ}\),
0.037 for \(V_M\),
0.031 for \(V_V\), 0.021 for \(V_{HH}\),
and 0.0044 for the hybrid \(V\)\@.
The hybrid form \(V\) is five times more
accurate than the
class-2 potentials (\ref{Morse}--\ref{HH})
and four orders of magnitude more accurate 
than the class-1 potentials (\ref{LJ}--\ref{Ha})\@.
\par
Can \(V(r)\) also represent 
weak noncovalent bonds?
Using Gnuplot, we fitted 
\(a\), \(b\), \(c\), and \(e\) in Eq.(\ref{phenpot})
to Aziz's accurate HFDID1 potential for 
Ar\(_2\)~\cite{Aziz1993} and HFD-B potential
for Kr\(_2\)~\cite{Aziz1986}
and set \(d\) equal
to their London-tail coefficients.
Figures~\ref{ar237}--\ref{kr2} 
show that for \(2 \le r \le 7\)\AA,
the hybrid form
(\ref{phenpot}) (solid, red) 
fits the Aziz potentials for both Ar\(_2\) and
for Kr\(_2\) (pluses, blue)\@.
The Lennard-Jones \(V_{LJ}\) curves
(\ref{LJ}) (dashes, green) 
matched at the minima
are too deep for \( r > 4.5 \)~\AA\ 
(Figs.~\ref{ar237} \& \ref{kr237})
and too hard for \( r < 3 \)~\AA\ 
(Figs.~\ref{ar2} \& \ref{kr2})\@.
%Being accurate at all relevant distances,
The potential \(V(r)\) of Eq.(\ref{phenpot}) 
represents weak noncovalent bonds
better than \(V_{LJ}\)
(and \(V_H\))\@.
\par
Does it matter that \(V_{LJ}(r)\)
fails to fit 
the Ar--Ar and Kr--Kr interactions?
To find out, we used \(V\) and  \(V_{LJ}\)
to compute the dimensionless second virial coefficient \(B_2/r_0^3\)
of Ar and Kr at room temperature (\(kT = 0.025 \) eV)\@.
Here \(r_0\) is the minimum of the potential, and
the second virial coefficient \(B_2\) is the integral
over all space
\beq
B_2(T) = - {\thalf} \, \int \! d^3r \, 
\left( e^{-\beta V(r)} -1 \right) 
\label{B_2}
\eeq
in which \(\beta = 1/(kT)\)\@.
The hybrid potential \(V\) 
fitted to the curves of Figs.~\ref{ar237} -- \ref{kr2}
gives \(B_2/r_0^3 = -0.499\) for Ar and 
\(-1.35\) for Kr,
which respectively differ from 
the experimental~\cite{Lide1994}
values of \(-0.552\) and \(-1.41\) by 9.6\% and 4.1\%\@.
The Lennard-Jones potential \(V_{LJ}\)
fitted to \(r_0\) and \(V(r_0)\)
gives \(B_2/r_0^3 = -0.899\) for Ar and \(-1.92\) for Kr
(errors of 63\% and 37\%);
it also wags a long-range tail with  
London coefficients \(2 \, |V(r_0)| \, r_0^6 \)
that are too large by 83\% for Ar and by 84\% for Kr\@.
If the parameters \(r_0\) and \(V(r_0)\)
in Eq.(\ref{LJ}) for \(V_{LJ}\)
are chosen to give the correct 
second virial coefficient \(B_2/r_0^3\)
for a range of temperatures~\cite{Hill1986},
then \(V(r_0)\) 
is too shallow
by 16 \% for Ar and 15\% for Kr,
and the London coefficients of the 
long-range tail are too large by 70\% for Ar and by 64\% for Kr\@.
With only two parameters,
Lennard-Jones fits are procrustean.
\par
What about additivity (\cite{Axilrod1943,Muto1943,Cole1988})?
When three or more atoms interact,
their potential energy is not
the sum of the three (or more) pair potentials.
Is the accuracy of the hybrid form important
in the liquid phase
where additivity is only approximate?
\par
To test whether the lack of complete additivity
in the liquid phase obscures the 
advantages of the hybrid form \(V\) over the
Lennard-Jones potential \(V_{LJ}\),
we used both to compute  
the heats of vaporization \(\Delta_{\mathrm{vap}}H\)
of Ar and Kr at their
boiling points at atmospheric pressure.
In our Monte Carlo simulations,
we imposed periodic boundary conditions 
to reduce finite-size effects.
Our Monte Carlo code is 
available at 
\href{http://bio.phys.unm.edu/latentHeat/}
{bio.phys.unm.edu/latentHeat}.
We used it
to compute the potential energy \(U\) per
atom in the liquid and gas phases.
The latent heat of vaporization 
\( \Delta_{\mathrm{vap}}H \)
is the difference between the potential
energies \(U_{\textrm{gas}}\) and \(U_{\textrm{liquid}}\)
plus the work done in expanding by \(\Delta V\)
against the pressure \(p\)
of the atmosphere,
\(\Delta_{\mathrm{vap}}H = U_{\textrm{gas}} \,
- \, U_{\textrm{liquid}} \, + \, p \, \Delta V \)\@.
\par
The hybrid form \(V\) 
fitted to the curves 
of Figs.~\ref{ar237} -- \ref{kr2} gave
\(\Delta_{\mathrm{vap}}H = 0.0694\) eV (per atom) for Ar
and 0.0982 eV for Kr,
which differ from 
the experimental values~\cite{Lide1994} of 0.0666
and 0.0941 eV by 4.2\% and 4.4\%\@.  % CRC
%and 0.0932 eV by 5.0\% and 6.0\%\@. % NIST
In equivalent Monte Carlo simulations,
the Lennard-Jones potential \(V_{LJ}\) fitted
to \(r_0\) and \(V(r_0)\) gave
and \(\Delta_{\mathrm{vap}}H = 0.0787\) eV for Ar 
and 0.111 eV for Kr
(errors of 18\% and 18\%)\@.
So the errors due to a lack of additivity
are of the order of 4\%,
while those due to the Lennard-Jones potential
are about 18\%\@.
Even in the liquid phase,
limited additivity is less of a problem
than the defects of the Lennard-Jones potential.
\par
For a wide class of atom pairs,
the hybrid form 
(\ref{phenpot})
can reliably represent the best
spectroscopically determined
potentials over all relevant distance scales. 
It also yields accurate second virial coefficients
and heats of vaporization.
Its simplicity recommends it as a teaching tool
and as a practical form for computation.
Given the differences between it
and the 
Lennard-Jones, harmonic, Morse,
Varnshi, and Hulburt-Hirschfelder potentials, 
it would be worthwhile
to examine the consequences of
these differences in Monte Carlo searches
for low-energy states of biomolecules
and in numerical simulations
of phase transitions and reactions
far from equilibrium.
\begin{acknowledgments}
Thanks to S.\ Valone for conversations 
and for RKR data;
to S.\ Atlas, C.\ Beckel, 
B.\ Brooks, J.\ Cohen, K.\ Dill, D.\ Harries, G.\ Herling,
M.\ Hodoscek, R.\ Pastor, 
R.\ Podgornik, W.\ Saslow, and C.\ Schwieters for advice.
P.~J.\ Steinbach kindly hosted KC\ at NIH,
where we used the Biowulf computers.
\end{acknowledgments}

\bibliography{chem,cs,physics,vdw}

\end{document}